\documentclass{JHEP} 
\usepackage{amssymb} 
\usepackage{amsfonts}
\newcommand{\beq}{\begin{equation}}
\newcommand{\eeq}{\end{equation}}
\newcommand{\be}{\begin{eqnarray}}
\newcommand{\ee}{\end{eqnarray}}

\newcommand{\dd}{{\mathrm d}}
\newcommand{\e}{{\mathrm e}}

\newcommand{\Leg}{\mathrm{P}}
\newcommand{\CC}{\mathbb{C}}
\newcommand{\In}{\mathrm{in}}
\newcommand{\Out}{\mathrm{out}}
\newcommand{\Res}{\mathrm{Res}~}
\newcommand{\half}{\frac{1}{2}}
\renewcommand{\Im}{\mathrm{Im}}
\renewcommand{\Re}{\mathrm{Re}}

\title{Decay Modes of Highly Excited String States and Kerr Black Holes}

\author{
Roberto Iengo and 
Jussi Kalkkinen\thanks{This work was supported 
by the European Union TMR program CT960045.} \\ 
SISSA, via Beirut 4, Trieste 34014, Italy \\
INFN, Sezione di Trieste \\ 
E-mail: {\tt iengo@sissa.it, kalkkine@sissa.it}
}

\abstract{We consider four-graviton scattering in Type II string 
theory on one-loop level in the large centre of mass energy $M^2$
limit. 
We extract from it an explicit integral
expression for the full string theory corrections to the imaginary part of the 
mass-shift and the lifetime of a massive state with the highest allowed spin 
$J=2M^2+2$. 
We find a decay rate that is  up to $\log M$ corrections of order $1$ in
string units, times $g_s^2$.  
We also find that the dominant decay mode corresponds to the 
emission of  light particles, whereas the decay into two massive or two
massless states is exponentially suppressed. We discuss the relation of 
our results to quantum gravity aspects of a Kerr Black Hole.}


\preprint{SISSA 79/2000/EP\\
{\tt hep-th/0008060}}

\begin{document}

\section{Introduction}

String theory provides a very nontrivial regularization
of ultraviolet phenomena in gravity. One of the many surprising features
of this regularization is the fact that increasing the mass of a state
does not necessarily lead to decreasing its lifetime; string theory
dynamics seems to know how to suppress the 
kinematically allowed phase space crucially. 
We calculate the decay width and the mass-shift of a massive string state
that displays this behaviour. 
As we are far outside the validity of a field theory analysis
we have to consider all string theory corrections
in the calculation at the same time, in order to see whether it really persists;
the field theory limit is not enough. 

We will perform the evaluation of the lifetime by means of a direct
computation: We  consider the one-loop contribution to the mass-shift of
the massive state and evaluate its imaginary part in 
the Type-IIA or -IIB superstring theory. We will obtain the relevant
amplitude by starting from the well known formula for the one-loop four
graviton scattering amplitude in superstring theory, and extracting from
it the residue of the singularity when the Mandelstam variable $s$
approaches the mass of the massive state. We have studied the case of the
(unique, nondegenerate) maximal angular momentum state, which we can select by
looking at the dependence on the scattering angle, fixing the normalization by
comparing it with the residue at the pole of the same four graviton amplitude
at tree-level. Since we work at one-loop level, we are considering the lifetime
as the massive state is allowed to decay into two states only; 
this is the only possibility at the first nontrivial order 
in the perturbative expansion.

The analysis is rather subtle, because the resulting amplitude is expressed
as an integral over tori: this boils down to integrating over 
a four real dimensional manifold 
that consists of both the torus surface and the fundamental domain of 
the complex modular parameter of the 
torus. This integral is formally divergent and, as usual in string
theory, it must be given a meaning by analytic continuation. It is
precisely this feature which gives rise to an imaginary part, as otherwise
the result would be real, as the integrand is real positive. 
This is also very similar to what happens in standard Feynman loops, for
instance in a convenient exponential parametric representation of the
propagators: the resulting integral is convergent in the Euclidean domain,
when the Mandelstam variable $s$ is space-like, and it is formally divergent 
if $s$ is time-like, which is relevant for the decay. The correct
result is obtained by analytic continuation.
In the case of string theory one cannot start from a naive off-shell
configuration, as there are consistency requirements, such as preserving
modular invariance, but the strategy is similar. We have first checked the
method by computing the lifetime for the decay of the massive state into
two massless string modes, which can be compared -- and it agrees -- with a
direct computation of the decay into two gravitons. This 
can be done by extracting the tree-level coupling of the massive state to
two gravitons from the tree-level four graviton amplitude.
The result is that this decay channel is exponentially suppressed for our
massive state (it goes like $\sim \exp{-\beta M^2}$ for $\beta>0$ a numerical constant).

The full investigation, including any two body channel, is based on the general
method of saddle point analysis, although we also have to include contributions 
from more singular loci than saddle points in the integral: It turns out 
that the dominant behaviour arises in regions where the 
integrand  has a conical singularity. Visualizing the behaviour of 
the integrand on sections of the manifold has required some
numerical work. 
Happily, the numerical analysis confirms that the relevant configurations always correspond to
points having particular geometrical and/or symmetrical meaning. In fact,
it turns out that the number of relevant configurations is rather
limited and that we could reliably follow the evolution of the integrand on
them.

The final result is that the lifetime of our massive state 
is of order $1$, up to logarithmic corrections.
It is interesting that also a physical picture emerges from our
analysis: The dominant decay mode is seen to 
be rather asymmetric, corresponding to 
a radiative-like process, where the very massive state decays into an
other very massive state emitting a light particle. The other
configurations, where the massive state splits into two states, which are
both massive or both light, are exponentially suppressed.        

In \cite{Gross:1988ar} the behaviour of the four graviton fixed angle
scattering was investigated. 
Their results are in qualitative agreement with ours in that
they, too, find exponentially 
suppressed decay widths for decays into light particles, 
gravitons in their case.
Also there the dominant contributions arise in 
geometrically significant points.  
In Ref.~\cite{Balasubramanian:1996xf} 
the decays of open string excitations of D-branes were studied, 
for polarizations either parallel or transverse to the brane; 
for transverse polarizations the leading decays also arise from 
the emission of soft light particles. 
Ref.~\cite{Sundborg:1989ai} deals with a similar problem to ours, 
though finding a somewhat different result.
Related issues were considered in Ref.~\cite{Damour:2000aw}.

The fact that the dominant decay mode is the emission of 
massless particles is reminiscent of Hawking radiation.
Also, it is expected that the collision of gravitons at very high energy
may lead to the formation of a gravitationally collapsed intermediate
state \cite{Amati:1987wq,Feinstein:2000ja,Bozza:2000vk}. 
We are thus lead to compare our results to black hole physics. 
The distinctive features of our state are
its high angular momentum and the fact that it is in the range of
validity of perturbative string theory. Therefore it should, somehow,  correspond
to an extrapolation of a Kerr black hole to the small string coupling region.
However, the radiation rate we obtain is slightly higher than that of the Hawking radiation of
a black hole. 
Indeed, the detailed analysis of  Section \ref{discussion}  
indicates that our state also extends beyond the 
horizon radius of the corresponding
classical Kerr solution, when the string coupling is in the perturbative
region. 
Requiring its characteristic length scale   to be of the same order as the
classical horizon turns out to be equivalent to requiring that the string coupling constant
 have the same -- large --  value as what results from the
Correspondence Principle \cite{Horowitz:1997nw} 
(see also \cite{GVFC,Bowick:1987km,'tHooft:1990fr,Susskind:1993ws}). 
We are thus lead to expect that the extrapolation of a classical Kerr black hole 
to the quantum perturbative regime corresponds to a very 
massive string state of large angular momentum.

\section{Formula for the lifetime}

We start by considering the one-loop four-graviton amplitude in the
Type IIA/B superstring  theory
\be
 A_1 = R^4 \int {\dd^2 \tau \over (\Im\tau )^2}\int \prod_{i=1}^3 
{\dd^2 z_i \over \Im\tau}
        \e^{-2\sum k_i \cdot k_jP(z_{ij})}          ~,       
\ee
where $R^4$ is a kinematical factor constructed from the tensor
$t_8$ \cite{Schwarz:1982jn} that contains the graviton polarizations.  
$z_i$ are the 
puncture insertion points of the graviton vertices, with
$z_{ij}=z_i-z_j$ and $z_4=0$,
to be integrated over the torus surface. 
$\tau$ is the torus modulus to be
integrated over the fundamental domain. $P(z_{ij})$ is the propagator of
the world-sheet scalar 
\be
P(z)&=&\ln\left|{\theta_1 (z|\tau )\over \theta_1^{'} (0|\tau )}\right|^2-2\pi {(\Im z)^2
\over \Im\tau}~.
\ee
For the  entering graviton momenta $k_i$ the Mandelstam invariants are
$s=2k_1\cdot k_2=4E^2$, $t=2k_1\cdot k_3=-2 E^2(1-\cos\theta)$, 
$u=2k_1\cdot k_4=-2E^2(1+\cos\theta)$
where $E$ and $\theta$ are the  graviton energy and the scattering
angle in the CM frame. We work in units of $\alpha^{'}$.

The amplitude $A_1$ has a double pole for $s\to N$ due to the propagator of a 
massive string
state with $M^2=N$, produced by the collision of the two incoming
gravitons entering into the loop (i.e.~the torus), and another similar
propagator emerging from the loop and decaying into the outgoing gravitons.
The residue of this double pole is proportional to $\Delta M^2$, 
the mass-shift of the massive state: $M^2 \to M^2 + \Delta M^2$. 
The imaginary part of $\Delta M^2$ gives the inverse lifetime of the state.  

In order to get $\Delta M^2$ we divide the double pole residue of $A_1$ by
the residue of the simple pole (for $s\to N$) of $A_0$, the tree-level
four-graviton scattering amplitude. In fact, 
\be
A_1 \to G_{\In}{1 \over s-N} \Delta M^2 {1 \over s-N} G_{\Out}~,
\ee
where $G_{\In(\Out)}$ represent the coupling of the state to the
incoming (resp.~outgoing) gravitons, and
\be
A_0 \to G_{\In}{1 \over s-N} G_{\Out}~.
\ee

The poles of $A_1$ for $s\to N$ occur as singularities of the integral over
$z_i$ for $z_{12} \to 0$ and $z_{34} \to 0$. The integrand behaves as
\be
\sim |z_{12}|^{-2s}|z_{34}|^{-2s} \cdot F(z_{12},z_{34},z)~,
\ee
where $z=(z_1+z_3)/2$. In order to get the poles we have to look at the
terms of the expansion of the remaining function $F$ that behave as
$|z_{12}|^{2(N-1)}|z_{34}|^{2(N-1)}$.
Let us now 
consider the string state with maximal angular momentum
$J=2N+2$ and $M^2=N$, and thus look for the maximal power of $\cos\theta$,
or else of $(t-u)$, in the residue. We find a unique term of that kind in 
the expansion of $F$, namely 
\be
F(z_{12},z_{34},z) 
&\sim&  (E^2 \cos\theta )^4 ~ (\epsilon_{1} \cdot \epsilon_{2}~ 
        \epsilon_{3}\cdot \epsilon_{4})~ \left({t-u \over 2}\right)^{2N-2} 
        {1\over (2N-2)!}~ \e^{2N~P(z)}    \cdot ~\nonumber \\
&    &    |z_{12}|^{-2N} |z_{34}|^{-2N} 
           \left( (f^{``}(z)+{\pi\over \Im\tau})z_{12}z_{34} \right.
                          \nonumber \\
&  &     + \left.(\bar f^{``}(z)+{\pi\over \Im\tau})   
         \bar z_{12}\bar z_{34}-{\pi\over \Im\tau}
         (z_{12}\bar z_{34}+\bar z_{12}z_{34}) \right)^{2N-2} ~,
\ee
where the term $(E^2 \cos\theta )^4 (\epsilon_{1}\cdot\epsilon_{2}~ \epsilon_{3}\cdot\epsilon_{4})$ 
arises from $R^4$, $\epsilon_i$ being the graviton polarization tensor, 
and $f^{``}=\partial_z^2 \theta(z|\tau)$.

The residue of the double pole, for the maximal $J$, turns out to be
\be
\Res A_1 
        & \sim &  (E^2\cos\theta )^4(\epsilon_{1}\cdot \epsilon_{2}
                      ~     \epsilon_{3}\cdot \epsilon_{4}) 
\frac{1}{(2N-2)!} \left({t-u \over 2}\right)^{2N-2}
             \nonumber  \\
        & &  \cdot~ \int {\dd^2\tau \over (\Im\tau )^2}\int {\dd^2 z \over \Im\tau}
            \e^{2NP(z)} ~  \sum_{n=0}^{N-1}{(2N-2)! \over (n!)^2(N-n-1)!^2} \nonumber \\
        & &  \cdot~ \left|f^{``}(z)+{\pi\over \Im\tau}\right|^{2n}
                                        \left({\pi\over \Im\tau}\right)^{2N-2n}~.
\ee
On the other hand, the residue at the pole for $s \to N$ 
of the tree-level amplitude for the maximal
$J$ is obtained from the Veneziano amplitude for the closed superstring
\be
A_0= R^4{\Gamma (-s)\Gamma (-t)\Gamma (-u)
         \over  \Gamma (1+s)\Gamma (1+t)\Gamma (1+u)}
\ee
and it is
\be
\Res A_0  &\sim&  \left({t-u \over 2}\right)^{2N-2}{1\over (N!)^2}
          (E^2\cos\theta )^4(\epsilon_{1}\cdot \epsilon_{2} \epsilon_{3}\cdot \epsilon_{4})~.
\ee
Finally, using  a standard formula for the Legendre polynomial, 
\be
\Delta M^2 &\sim&   N^2
          \int {\dd^2\tau \over (\Im\tau )^2}\int {\dd^2 z  \over \Im\tau}
          \e^{2NP(z)}~ ({\pi\over \Im\tau})^{2N}~   \nonumber \\
& &        \cdot ~    (Q-1)^{N-1}\Leg_{N-1}({Q+1\over Q-1}) \label{starting}     ~,   
\ee
where $\Leg_n$ is the Legendre polynomial of degree $n$ and 
$Q=|f^{``}(z)\Im\tau/\pi +1|^2$. The sign $\sim$ indicates here, and elsewhere in this article, 
that we have left out an $N$-independent normalization.
As in this expression  the combinations  $Q$ and $\pi\exp(P)/ \Im\tau$ are separately 
modular invariant the mass-shift $\Delta M^2$ is well defined.


The inverse lifetime of the massive state is given by the imaginary part of
$\Delta M^2$, divided by $M=\sqrt N$
\be
\Gamma &=& {\Im \Delta M^2 \over \sqrt N}~.
\ee
Note that $\Delta M^2$ is expressed, formally, as an integral of a real
quantity, but this integral is actually divergent. The imaginary part comes
from the fact that this expression has a meaning in the sense of an analytic
continuation.

\section{Decay rate into gravitons}

As a first investigation of the Physics of the decay of the massive string
mode we compute its lifetime for decaying into two gravitons. This
computation can be done in two ways: 

\begin{itemize}
\item[] {\em First}, by extracting the coupling of
the massive state to the gravitons by looking at the pole of the tree-level
amplitude $A_0$ and then performing the integral over the phase space, or,
\item[] {\em Second}, by extracting the contribution of the two graviton channel
from the formula for the inverse lifetime we derived, starting by the double
pole residue of the one-loop amplitude $A_1$. This can be done by looking
at the dominant contribution in the pinching limit $\Im\tau\to\infty$ of the
torus, corresponding to massless states circulating into the loop.
\end{itemize}
In the first strategy we start by recalling that
\be
\Res A_0 &\sim& (2p^2\cos\theta )^{J}~ {1\over (N!)^2}~
          (\epsilon_{1}\cdot \epsilon_{2}\epsilon_{3}\cdot \epsilon_{4})~,
\ee
where the CM square space momentum is $p^2=s/4$. As said, $J=2N+2$ and we
have taken the maximal power of $\cos\theta$. 
On the other hand, the
Feynman graph describing two incoming gravitons forming a massive state of
angular momentum $J$ further emitting two outgoing gravitons would be 
(looking at the maximal power of $\cos\theta$)
\be
A_0 & \sim & (\epsilon_{1}\cdot \epsilon_{2}\epsilon_{3}\cdot \epsilon_{4})~ 
             {g_J^2 \over s-M^2} ~
\sum_V p^{r_1}\cdots  p^{r_J} 
            V_{ r_1\cdots r_J}~ V_{ s_1\cdots s_J} p'^{s_1}\cdots  p'^{s_J}
             \cdot\nonumber \\
     &\sim& (\epsilon_{1}\cdot \epsilon_{2}\epsilon_{3}\cdot
             \epsilon_{4})~
             {g_J^2 \over s-M^2}~ (p^2\cos\theta )^J ~,
\ee
where $V_{ r_1\cdots r_J}$ is the (symmetric and traceless) polarization
tensor of the massive state, and the indices $r_i$ run over the $d=9$ space dimensions. 
By comparing the pole residue we get that (up to an $N$-independent
coupling constant):
\be
g^2_J &=& {2^{2N+2} \over  (N!)^2}~.
\ee
Now we can directly compute the inverse lifetime for the decay into two
gravitons. It is (up to $N$-independent numerical constants)
\be
\Gamma_{\mathrm{2grav}} & \sim & {1\over \sqrt N}g^2_J~ \int {\dd^d 
p \over p^2}~ \delta (p-{\sqrt s\over 2})~
                     V_{r_1\cdots r_J}p^{r_1}\cdots p^{r_J}
                     p^{s_1}\cdots p^{s_J}V_{ s_1\cdots s_J} \nonumber \\
               &\sim&  {1\over \sqrt{N}} g^2_J~ p^{d-3}  p^{2J}J^{-{d-1\over 2}}~
                          {2^JJ!^2\over (2J)!} ~.
\ee
We have assumed the normalization $||V|| =1$ and made use of the formula
\be
\int \dd^{d} \hat{p}~ \delta (\hat{p}^2 -1)
                 V_{r_1\cdots r_J}\hat{p}^{r_1}\cdots \hat{p}^{r_J}
                 \hat{p}^{s_1}\cdots\hat{p}^{s_J}W_{ s_1\cdots s_J} 
                  \sim V\cdot W ~ J^{-{d-1 \over 2}}~{2^JJ!^2\over (2J)!}~,
\ee
which holds for symmetric and traceless $V$ and $W$. The angular
integration for large $J$ gives the factor $J^{-{d-1 \over 2}}$.           

Notice that the dependence on the dimensionality $d$ disappears in the
final result.
By using the Stirling formula for large $N$ and $J=2N+2$ and putting 
$p^2=N/4$, we get that the
inverse lifetime for the decay into two gravitons is exponentially small:
\be
\Gamma_{\mathrm{2grav}} &\sim&  ({\e\over 4})^{2N}~.
\ee
This is already a surprising result, because one would have expected the
massive state to decay very rapidly due to the large phase-space. Instead,
the indication is that the very high mass, very high angular momentum string states
might have a long lifetime. We
will see that that this is indeed true also when summing over all the
two-body channels: The decay into two states of comparable mass
is in general exponentially suppressed, 
although the dominant contribution coming from the decay into a massless
and another very massive state is of order unity $\sim N^0$, times logarithmic corrections.

For the moment we continue with the computation of the lifetime for decay
into two massless particles by using now the second method, based on the 
one-loop (torus) amplitude and looking at the pinching limit of the torus.

We write $\tau =\tau_1 +i\tau_2$ and $z=x+y\tau$. Thus $x,y$ vary between
$0$ and $1$ whereas $\tau$ spans the fundamental domain. The relevant pinching limit
corresponds to $\tau_2\to\infty$ with $y\not= 0,1$. In this limit we have,
putting $y=1/2+\eta$,
\be
\e^{2NP(z)} &\to& {\e^{N\pi\tau_2}\over (4\pi^2 )^{2N}}\e^{-4N\pi\tau_2 \eta^2}~.
\ee
Further, in this limit $Q\to 1$ and we get
\be
\lim_{Q\to 1} (Q-1)^{N-1}P_{N-1}({Q+1\over Q-1})={(2N-2)!\over (N-1)!^2}~.
\ee
In order to keep only the contribution of the two-massless particle
channel, we replace the integrand with its pinching limit expression and
get
\be
 \Gamma_{\mathrm{2 massless}}  &\sim&  N^{3/2} {(2N-2)!\over (N-1)!^2}~ 
         \Im\int {d^2\tau \over \tau_2^5}{\e^{N\pi\tau_2}\over (4\pi^2 )^{2N}}
                          ({\pi\over\tau_2})^{2N-2}~  
                        \int\dd\eta~ \tau_2 \e^{-4N\pi\tau_2 \eta^2}~. \label{pinch}
\ee
After doing the Gaussian integral in $\eta$, we have to get the imaginary part
of the integral over $\tau$, which is expected from the integral over $\tau_2$
extending up to infinity and formally divergent.
We have thus to consider 
\be
\Im\int \dd\tau_2 \tau_2^{-2N-2-1/2}\e^{N\pi\tau_2} \sim 
                {(N\pi )^{2N+1+1/2} \over \Gamma(2N+2+1/2)} ~.
\ee
This result is obtained by analytic continuation. A quick way of getting it
is to look for the saddle point of the integrand: We find a minimum at 
$\tau_2 =({4 \over \pi})^2$ and thus expanding around it we get an inverted
Gaussian which integrates to an imaginary result.
By putting the various factors together we finally get
\be
\Gamma_{\mathrm{2 massless}} &\sim &   \left({\e\over 4}\right)^{2N}~,
\ee
that is, up to a numerical factor, the same as $\Gamma_{\mathrm{2grav}}$ computed
directly by Feynman rules. Evidently the other possible massless channels
beside the two gravitons give the same large $N$ behaviour.

\section{Decay rate into string states}

We shall now calculate the lifetime of the massive state when 
it is allowed to decay also to massive string excitations in addition to
the massless gravitons considered above.
This amounts, in principle, to performing the integral (\ref{starting}) 
over the full fundamental domain and the torus.

\subsection{Saddle point analysis}

As we are only interested in the large $N$ limit of the amplitude, 
the leading contributions should arise, at least if the integral were convergent,  
from where the integrand reaches its maximum. In the present case 
we shall 
have to deal with the  difficulty that the integral is actually divergent:  
As we saw above, we could regularize it by analytically continuing one of the
variables of integration into the complex plane; The correct leading 
contribution then came from near the point where 
the original integrand reached its minimum.

In this context, saddle point analysis is actually not just one of the
many ways to regularize a divergent integral. The amplitudes here, 
as they are in field theory in general, are generically divergent: 
What actually is physically significant, 
is the behaviour of the integrals near the extrema 
of the integrands after an analytic continuation (Wick rotation) 
in the variables of integration. 
Hence, we should find the extrema of the integrand and to consider
contributions coming from a Gaussian analysis near them. 
The integration contour along the parameters of the unstable directions should then be 
deformed in the complex plane to coincide with the imaginary axis in a local neighbourhood of the saddle point, so that the Gaussian integral could finally be calculated. 

We should also consider separately all the 
singularities, orbifold 
points, and any other special points 
that one might run into in the integration 
region, as they will play the role of 
end points of integration.
In our case, the integration manifold has no true
boundaries, since it is described by two complex variables: the coordinate $z$
over the torus surface, which has no boundary, and the torus modulus
parameter $\tau$, which runs over the fundamental domain folded
into itself by modular transformations (we will recall below the most
relevant facts).  
For the 
convergent case the reason for this is clear: one is supposed 
to calculate the contribution coming from these points and simply 
to find out if it might be
the leading one or not. For the divergent case it is useful to think 
of these special points as kind of 
topological defects in the integration range. 

The formula (\ref{starting}) for the mass-shift can be conveniently cast in the form 
\be
\Delta M^2 &\sim&  N^2
          \int {\dd^2\tau \over (\Im\tau )^2}\int {\dd^2 z  \over \Im\tau} ~
          \left(\e^{P(z)}~ {\pi\over \Im\tau}\right)^2 ~
          t(z|\tau)^{2N-2}~g_N(\sqrt{Q}) \label{simple}~,
\ee
where 
\be
t(z|\tau) &=& \e^{P(z)}~ {\pi\over \Im\tau} (\sqrt{Q}+1)\\
g_N(x) &=& \left({x-1 \over x+1}\right)^{N-1}\Leg_{N-1}({x^2+1\over x^2-1}) ~.
\ee
The function $g_N$ behaves very mildly in the large $N$ limit: 
On the positive real axis it is bounded between $1$ and 
$1/\sqrt{\pi (N-1)}$; in the large $N$  limit, for fixed $Q \neq 0$, it  actually  approaches 
\cite{szego} the value $\sim 1/\sqrt N$ except
for $Q$ very near to $0$. 
 By using the integral representation
of the Legendre polynomial, it is seen that this also holds for 
$Q$ growing as a power of $\log N$ (up to $\log N$ corrections).
We will see that this is the region where we get the maximal contribution.  
In our analysis we can hence 
safely approximate it with this value, 
and concentrate below on the function $t(z|\tau)$.
The integral behaves therefore at $N\longrightarrow \infty$ as 
\be
\Delta M^2 & \sim &  {N^{3/2} \over \sqrt{4 \pi}  }~
          \int {\dd^2\tau \over (\Im~\tau )^2}\int {\dd^2 z  \over \Im~\tau} ~         
          (Q^{3/4} + Q^{1/4})^{-1}~ 
          t(z|\tau)^{2N}  \label{szintegral}~.
\ee

\subsection{Critical Points}

We can expect {\em a priori} that, due to the symmetry of the problem,
the dominant contribution to the integral and, in particular, to its
imaginary part comes from the neighbourhood of points that have a
geometrical meaning. We will see that this is indeed the case. Also, it is
important to keep in mind the physical meaning of the coordinate $z$:
the double pole residue of the four graviton torus amplitude represents
the amplitude of two vertices (of the massive state) inserted on the torus
surface at the point $z$ and at the origin respectively.  

Let us review some main features. The integrand $t$, and also $Q$ as well,
are symmetric for $z \to -z$ and their first derivatives in $z$ vanish for
the points $z =  1/2$, $z= \tau/2$ and $z=1/2 + \tau/2$.
These points are related using the modular transformations 
\be
T: \tau & \mapsto& -1/\tau \\
S: \tau & \mapsto& \tau + 1 
\ee
in such away that $T$ interchanges $z = 1/2$ and $z=\tau/2$ but maps  
$z=1/2 + \tau/2$ back to itself, whereas
$S$ interchanges $z = 1/2+\tau/2$ and $z=\tau/2$ but maps  $z=1/2$ back to
itself. The point $z=0$ is invariant under the modular transformations.

Both $t$ and $Q$ are modular invariant and the complete integrand,
including the measure, is also modular invariant. Therefore the
integration over the modulus $\tau$ is restricted as usual to the
fundamental domain: $|\tau_1 | \leq 1/2$ {\em and} 
$|\tau | \geq 1$. 
From now on we denote $\tau_1\equiv \Re~\tau$ and $\tau_2\equiv \Im~\tau$. 
 
Modular transformations relate, in general, points outside
the fundamental domain to points inside it, but the two lines $\tau_1 =\pm 1/2$
are identified by the transformation $S$, whereas the border segment
$|\tau | = 1,~~\tau_1 >0$ is identified with $|\tau | = 1,~~\tau_1 <0$ 
by the transformation $T$.
Thus there are  two orbifold points of the fundamental domain  
that are mapped back to themselves under modular transformations: These are
$\tau=i$  and $\tau = 1/2 + i\sqrt{3}/2$. This last point is invariant
under the full modular group and, there, all the three points 
$z =  1/2$, $z= \tau/2$ and $z=1/2 + \tau/2$ are identified.

Of course, another special point is the pinching limit of the torus
$\tau \to \infty$. In this limit the behaviour of the integrand depends
on whether $z$ remains fixed or not: 
\begin{itemize}
\item[1)]
For $z = 1/2$, we have $t \to 1$ and $Q \to \infty$. In this case the
integral is convergent.
\item[2)]
For both $z = 1/2+\tau/2$ and $\tau/2$ we have   
\be
t \to \e^{{\pi\tau_2 \over 2}} {2\pi \over 4\pi^2 \tau_2}
\ee 
and $Q \to 1$.
In this case the integral is divergent.
\end{itemize}
Since the integrand is real positive, we can only expect an imaginary part
from the fact that the integral, being formally divergent, has to be
defined by analytic continuation. One would thus be tempted to exclude the
point $z = 1/2$, but since it is related to the other points by modular
transformations it, too,  can -- and will -- play a role.

As for the point $z=0$, there we have $t=1$, for any $\tau$, and   
the integral is convergent, and thus it does not give rise to an imaginary part.
This point therefore contributes only a power dependence in $N$,  and only to the real part. 
Indeed it represents the pinching limit where the relative distance of
two vertices of the massive state coalesce: It 
corresponds to a tadpole correction to the massive propagator from which no
imaginary part is expected.  

It will be convenient to analyse the integral by choosing
to perform it over some geometrically 
meaningful modular invariant variable. A good part of the
investigation has been carried out numerically, by using the software 
{\em Mathematica} to visualize the shape of the integrand on various one-
and two-dimensional sections. Actually this numerical insight confirms the
expectation that the relevant contribution comes from the above discussed
critical points.

\subsection{Modular invariant analysis}
\label{fullcalculation}
\label{4.3}

Let us recall  that  $Q=(|w|~ \Im~\tau/\pi )^2$, where 
\be
w & =&\partial^2_z\theta_1 (z|\tau ) +\frac{\pi}{\Im~\tau}~,
\ee
is a modular invariant quantity. 
In order to reveal its  geometrical meaning  we  observe that
\be
w(z) &=& \wp(z) + 2 \zeta(\half) - \frac{\pi}{\Im(\tau)}~,
\ee
where $\wp$ and $\zeta$ are the respective Weierstrass functions. 
It is instructive to consider the integration over the 
complex variable $w$ instead of $z$.
Then the measure becomes
\be
\dd^2 z &=& \frac{\dd w}{u} \wedge \frac{\dd \bar{w}}{\bar{u}}~,
\ee
where $w$ and $u = \wp'(z)$ satisfy the equation of the torus 
\be
u^2 &=& w^3 +g_2(\tau) w + g_3(\tau)
\ee   
in $[w,u,1] \in \CC P^2$. When embedding the torus in the complex projective space, 
the new coordinate $w$ is therefore one of 
the projective coordinates, and the new integration measure is still the standard volume measure.
This change of coordinates is invertible except exactly at the branching points 
$z=1/2,\tau/2$ and $1/2+\tau/2$ where $\wp'$ vanishes. 

The exponential divergence of the integrand of Eq.~(\ref{szintegral}) 
for $\tau_2 \longrightarrow \infty$ 
appears only for $Q=1$. This is because, putting $z=x+y\tau$,
the integrand  diverges only for $y\not= 0$ and in this case 
$\partial^2_z\theta_1 (z|\tau) \longrightarrow 0$.
It might therefore be useful to consider the integral for a fixed value of 
$Q$, i.e.~integrating over $Q$ last ($Q$ can take all the values from $0$
to $\infty$). As far as the imaginary part of the mass shift is concerned, we can write
the integral (\ref{szintegral}) in the form 
\be
\Delta M^2 
& \sim &   N^{3/2}~ \int \dd Q~ \Big(~
          \int {\dd^2\tau \over (\Im\tau )^3}~ \int \dd \varphi~ 
          t^{2N} \Big) ~,
\ee
where $\arg w = 2 \pi \varphi$. This rearrangement is allowed because $Q$ 
is a modular invariant quantity. 
$\varphi$ is not invariant, instead, but gets shifted in modular
transformations. We observe that 
$\varphi \longrightarrow 0$ for $\tau_2 \longrightarrow \infty$.  

Let us briefly state our procedure: 
\begin{itemize}
\item[1)] We shall first identify where the dominant contributions arise for fixed
 $Q$ and $\tau_2$. 
\item[2)] We note that the imaginary unit appears, again, out of the integrations
over $\tau_2$. 
\item[3)] We find the imaginary part for all values of $Q$ separately, and notice that,
for different ranges of $Q$, the dominant contribution comes from different loci of the torus, 
thus corresponding to different physical processes. 
\item[4)] Finally, we notice that among these  contributions the dominant one 
arises in a particular limit $\tau_2 \longrightarrow \infty$, and we deal, in detail, with the 
subtleties of taking the limit.
\end{itemize}

We are now in the position to go into more detail: 
\begin{itemize}

\item[1)] We find the dominant contribution for fixed $Q$ and fixed $\tau_2$. This 
amounts to finding the maximum of the integrand $t$ as a 
function of $\tau_1$ and $\varphi$. For large values 
of $\tau_2$ and $\sqrt{Q} < 1.6259 $ ($\sqrt{Q} > 1.6259 $) 
this always appears at $\varphi = 0$ (resp.~at $\varphi = \pm 1/2$).  
Decreasing $\tau_2$ this maximum decreases monotonically, until, finally, 
another maximum in the $(\tau_1,\varphi)$-plane takes over.

\item[2)] When $\sqrt{Q} < 1.6259$ this dominant maximum on the  
$(\tau_1,\varphi)$-plane turns out to correspond 
to the point $z=1/2+\tau/2$ on the torus. This maximum is actually a conical
singularity: The first derivatives in $\tau_1$ and $\varphi$ are
discontinuous and do not vanish in this point. Following this maximum as a function of 
$\tau_2$ (for fixed $Q$), one finds that the value of the function fixed in the point corresponding 
to a {\em maximum} in   $\tau_1$ and $\varphi$,
goes through a {\em minimum} in $\tau_2$ at the boundary of the fundamental domain $|\tau|=1$. 
The fact that we found a minimum as
a function of $\tau_2$ relies on the fact that when we move over the boundary of the
fundamental domain at $|\tau|=1$ we can map back to the fundamental domain using the modular
transformation $T: \tau \mapsto -1/\tau$ that maps the point $z=1/2+\tau/2$
back to itself.

This minimum is smooth: The first derivative in $\tau_2$ is zero, and
we can therefore evaluate the inverted Gaussian integral that produces the 
imaginary part. We have thus found a ``saddle point'' contribution (for
fixed $Q$) to the imaginary part of the mass-shift. This point is  actually a conical
maximum on the $(\tau_1,\varphi)$-plane: integrating over $\delta\tau_1 ,\delta\varphi$ and $\delta\tau_2$
around it would therefore contribute a 
factor proportional to $i N^{-5/2}$. In this region
the integrand $t$ is less than $1$, corresponding therefore to an
imaginary part exponentially suppressed for large $M^2=N$.
 
\item[3)] This procedure can be performed for all values of $Q$ separately, and we have 
to pick the dominant contribution. When decreasing $Q$, the integrand $t$ at 
the saddle point decreases, and reaches its minimum at $Q=0$, where it
corresponds to $\tau =i$, which is a fixed orbifold point for $z=1/2+\tau/2$. 
When increasing $Q$ the value of the integrand $t$ at the saddle point increases. 
For $\sqrt{Q} =1.6259$ the
location of the saddle point in the $\tau$-plane reaches the corner of the
fundamental domain $\tau =1/2\pm i\sqrt{3}/2$ where the three points on the
torus $z=1/2+\tau /2$,$z=\tau /2$ and  $z=1/2$ are identified by modular transformations.  
When $\sqrt{Q}$ is increased above  $1.6259$, 
this saddle point (corresponding to  $z=1/2+\tau /2$) 
moves out of the fundamental domain (i.e. ~$|\tau_1| >1/2$), and local 
maxima -- still conical singularities on the  $(\tau_1,\varphi)$-plane --   
corresponding to the points  $z=1/2$ and $z=\tau/2$ move in. 
These points are interchanged under the modular transformation $T$. 

There is no minimum in $\tau_2$ inside the fundamental domain for the
conical point corresponding to  $z=\tau/2$, whereas there is a smooth
minimum in $\tau_2$ inside the fundamental domain for $z=1/2$. This minimum
occurs for $|\tau_1 |=1/2$. Thus, when $\sqrt{Q}$ passes through $1.6259$, 
the ``saddle point''
corresponding to $z=1/2+\tau /2$ is replaced continuously with the ``saddle point''
corresponding to $z=1/2$. We still get a factor $iN^{-5/2}$ from
integrating over the variations around it, but the integrand $t$ increases
with increasing $Q$ and ultimately  
$t \longrightarrow 1$ for $Q\longrightarrow\infty$. 
This is easy to check analytically, since for 
$\tau_2\longrightarrow\infty$ we have that, for $z=1/2$,
$\sqrt{Q}$ is proportional to $\tau_2$ and
$\theta_1 (1/2|\tau )/\theta^{'}_1 (0|\tau )\longrightarrow 1/\pi$.

\item[4)] The shape of the integrand flattens in the limit 
as $t \sim 1-O(\e^{-2\pi\tau_2})$. 
Therefore $t^{2N}$ goes down exponentially with $N$ until $\tau_2$ reaches the region 
$\tau_2 \sim\log N/2\pi$, where $t^{2N}$ begins to be of order $1$. Since the integral is
convergent for $t\sim 1$, it is suppressed for  $\tau_2 \gg \log N/2\pi$, 
and the conclusion is that the main contribution comes actually
from $\tau_2 \sim\log N/2\pi$.

Thus, finally, this region dominates since for $t \longrightarrow 1$ there
is no exponential suppression of the imaginary part. Rather, the imaginary
part will behave as a power of $N$. 

In order to find that power we have to
have a closer look at the region where $\tau_2\to\infty$ and $z=1/2+\zeta$
with $\zeta$ small. 
In this region we can use an approximation for the theta functions
\be
{\theta_1 (z|\tau ) \over \theta_1^{'} (0|\tau )}
={\cos(\pi\zeta )\over \pi} \left(1+4\cos^2(\pi\zeta ) \e^{i2\pi\tau}\right)
\ee
up to terms further exponentially suppressed for $\exp{(-2\pi\tau_2)}\to 0$.
In the same approximation we get
\be
\sqrt Q{\pi\over\tau_2}\e^{i2\pi\phi}=
-{\pi^2\over\cos^2(\pi\zeta )}-8\pi^2 \cos(2\pi\zeta)~ \e^{i2\pi\tau}
+{\pi\over\tau_2}~.
\ee
This relation determines the values  $\tau_1^*$ and $\phi^*$ at $\zeta =0$,
as functions of $\sqrt Q$ and $\tau_2$, in the region where both go to infinity. 
In particular, for $\tau_2\to\infty$ we get $\phi^*\to 1/2$. 
We want to investigate the behaviour of our integrand $t(\zeta ,\tau )$ 
as a function of $\tau_1$ and $\phi$ around $\tau_1^*$ and $\phi^*$
keeping fixed, and large,  $\sqrt Q$ and $\tau_2$. We have thus to
re-express the complex variable $\zeta$ as a function of 
$\delta\tau_1\equiv\tau_1 -\tau_1^*$ and $\delta\phi\equiv\phi -\phi^*$. 
From the previous equation we find (neglecting further suppressed terms):
\be
\sin^2 (\pi\zeta )&=& -i2\pi \left(\delta\phi~ {\pi\sqrt Q\over\tau_2}\e^{i2\pi\phi^*}+
                            \delta\tau_1~ 8\pi^2\e^{i2\pi\tau_1^*}\e^{-2\pi\tau_2}\right)~.
\ee
We notice that the dependence on $\delta\tau_1$ is exponentially
suppressed everywhere, both here and in the expression for the integration 
function $t(\zeta ,\tau )$.
Therefore, the  direction along $\tau_1$ is almost flat:
$t^{2N}\sim \exp{(-cNe^{-2\pi\tau_2}|\delta\tau_1 |)}$ and thus, in the
relevant region $\tau_2 \sim\log N/2\pi$, we do not get additional powers of
$N$ from the integration over $\delta\tau_1$.

Along the other direction we will still have conical behaviour, since
we get   
\be
\zeta^2=i \delta\phi~ {2\sqrt Q\over\tau_2}
\ee
from which we get that $\Re(\zeta^2)=0$ and 
$(\Im\zeta )^2=|\delta\phi|{\sqrt Q\over\tau_2}$. Substituting in the
approximate expression for the theta functions we get 
\be
t(\zeta ,\tau )\sim {\sqrt Q\over\pi\tau_2}\exp{\left( -{2\pi\sqrt Q\over\tau_2^2}|\delta\phi|\right)}~.
\label{4.17}
\ee
Therefore, integrating $t(\zeta ,\tau )^{2N}$ over $\delta\phi$ we get a
factor $1/N$.

The imaginary part of the result comes from the inverted Gaussian 
integration on $\delta\tau_2$ around the minimum in $\tau_2$ of $t^{2N}$ (as we
said, for large $Q$ and $z=1/2$, this minimum occurs at the (large) value of
$\tau_2$ corresponding to  $\phi =\tau_1= 1/2$). Also in this case, we do
not get additional powers of $N$ since this minimum is shallow:
$t^{2N}\sim \exp{(+cN\e^{-2\pi\tau_2}{(\delta\tau_2)^2})}$.  
 
\end{itemize}

\subsection{Conclusions}
\label{4.4}

Finally, inserting in
Eq.~(\ref{szintegral}) the contribution of this ``saddle point'', we get an imaginary part
of $\Delta M^2$ proportional to $N^{1/2}$, times some possible negative powers
of $\log N$ coming from the final integration over $Q$, whose precise value
depends on the details of the behaviour in $Q$ of the slopes of the saddle
point.

It is interesting to note that although restricting {\em ab initio} the
integration to the point  $z=1/2$ we would not expect any imaginary part in
the result, since for $\tau_2\longrightarrow\infty$ we would get in this
case a convergent integral of a real positive quantity, still this point on the torus
enters into the game as, in a sense,  the analytic continuation of the point
$z=1/2+\tau/2$, for which the integral is divergent and the imaginary part is expected.

In conclusion, we have found that the lifetime of the massive state of maximal
angular momentum behaves in the limit of large mass as a constant 
times  $\log N$ corrections. 
The dominant contribution comes from the neighbourhood of the
point $z=1/2$: This means that the two vertex insertions of the massive
state on the torus remain fixed at fixed positions on the torus 
even in the limit $\tau_2\longrightarrow\infty$
where the torus degenerates. Therefore, the dominant configuration
corresponds to a very long, and thus thin, handle attached to the two vertices, which 
are joined also by the finite part of the torus. 
The dominant string states on the long handle are the light ones. Therefore
the physical conclusion is that the dominant two-body decay mode of this
state is asymmetrical: It resembles a radiative
process where the very massive state decays into another very massive state
of lower mass by the emission of some massless  states.
Instead the fission-like processes, where the very massive state would split
into two more symmetrical fragments, maybe also of very small or even zero
mass (as we have explicitly computed in Section \ref{fullcalculation}), 
is exponentially suppressed.

\section{Discussion}
\label{discussion}

The result that the massive state decays by emitting low-mass particles 
suggests that it might be possible to interpret it as a black 
hole evaporating through Hawking radiation.
After a  charged black hole has stopped evaporating, the object that  
remains is believed to be a string theory bound state \cite{GVFC,Bowick:1987km,'tHooft:1990fr,Susskind:1993ws}. 
The evidence
for this has mainly come from comparing the 
geometric Beckenstein--Hawking entropy to the
string theory degeneracy of various D-brane configurations, and finding perfect agreement 
\cite{Strominger:1996sh}. Indeed, according to the Correspondence Principle 
of Horowitz and Polchinski \cite{Horowitz:1997nw}, when the size of the horizon drops below
the size of the string the typical black hole state becomes a typical string theory bound state.
It is therefore interesting to 
compare the massive state to a black hole at correspondence radius:

A massive state with mass $M_0 \sim \sqrt{N}$ can decay into a massive
state  with $M_1 \sim \sqrt{N-n}$ ($N$ large, $n$ finite) by emitting a
massless particle of energy $\sim 1/\sqrt{N}$. 
As the probability per unit time 
of this happening behaves as $\Gamma \sim 1$, we get the radiation rate
\be
\frac{\dd E}{\dd t} & \sim & \frac{1}{M} \label{life}~.
\ee
Note, however, that though it is clear that the dominantly emitted particles are massless,  
it does not follow directly from the above analysis that they were also soft. 
However, a more detailed comparison with field theory reported in the Appendix indicates 
that the dominant radiation is soft, although one gets actually a bound on $n$, 
that is $n/\sqrt{N} \longrightarrow 0$. One can also analyse the decay to a massless
state by a direct computation: a preliminary investigation indicates that the radiation is soft, with $n$ 
finite. We hope to return to this point in a future publication. 

The length scale  $R_J$ pertinent to the particular string theory state we are 
studying $|\Psi_J \rangle$,  for the maximal $J$, 
can be estimated by calculating the width of the state 
 $\langle \Psi_J |X^2|\Psi_J \rangle = R_J^2 \sim M^2$. 
The Correspondence Principle relates the string theory state 
to a black hole when the length scales are comparable; the 
radius of the corresponding
Kerr black hole \cite{Horowitz:1996tm} would therefore also be $\sim M$ and 
the area $A \sim M^{d-1}$. The Hawking 
radiation this black hole emits is that of a black body in 
temperature 
$T \sim M^{-1}$  \cite{Horowitz:1996tm},  and the
radiation rate is 
\be
\frac{\dd E}{\dd t} &\sim& A T^{d+1} \sim \frac{1}{M^2} ~, \label{res}
\ee
where $d$ is the number of space dimensions. 
The massive state radiates,
therefore, with a stronger intensity (\ref{life})
than a corresponding black hole. 
However,
if one were to identify the massive state with a black hole, 
string theory would have to be strongly coupled, as we shall presently see, 
making a straight forward comparison of the radiation rates difficult.

The ADM mass is related to the  radius of a black hole  
through $M_{\mathrm{BH}} \sim R_{\mathrm{BH}}^{d-2}~ g_{\mathrm{s}}^{-2}$, where $g_{\mathrm{s}}$ 
is the string coupling. Recall that at correspondence 
the length scale of the black hole is of the same order as the length scale of the quantum state 
$R_{\mathrm{BH}} \sim R_{J}$ and that in our system $R_J \sim M$.
Therefore, the string coupling where the ADM mass and the mass of the string theory state grow 
in the same way is
\be
g_{\mathrm{s}}^{2} &\sim& N^{\frac{d-3}{2}} \label{rad}~.
\ee
This means that we cannot reach the correspondence radius and simultaneously maintain 
the validity of string perturbation theory, as the calculation is performed in $d>3$. 
Therefore, the massive state cannot be directly thought of as a black hole but is, really, 
to be described as a quantum mechanical state. 

In $d+1\leq 5$ there is an upper bound for the angular momentum 
of a black hole with given mass that arises 
from  requiring that there be 
no naked singularities. As was noted in \cite{Myers:1986un,Horowitz:1996tm}, in  $d+1> 5$ 
there is no such bound, and all Kerr solutions posses a protecting horizon. 
It is, however, amusing to note that from the simple  requirement that 
the angular momentum arise from 
a mass distribution still inside the horizon, 
we get a bound $J < R_{\mathrm{BH}} M$, which reduces to an inequality
\be
g_{\mathrm{s}}^{2} &\gtrsim& N^{\frac{d-3}{2}}~.
\ee
This bound would be saturated at the correspondence radius (\ref{rad}): it seems, 
therefore, that in weak coupling, where we perform our calculations, 
a black hole interpretation is not feasible; On the other hand, 
when the black hole interpretation is adequate, the pertinent string theory is necessarily 
strongly coupled. 
Therefore, one could expect that varying the string coupling the system
undergoes
a phase transition, when the correspondence radius is reached.
This suggests that the descriptions in terms of a quantum state
and  in terms of a black hole are {\em complementary}, rather than equivalent.

\acknowledgments
We would like to acknowledge collaboration with
Daniele Amati in the early stages of this work, and to thank him for
proposing the study of excited string states at one-loop as a tool
for investigating quantum gravity effects. We thank also Matteo Bertolini 
and Kristian Jenssen for discussions.

\appendix
\section{Appendix}

In order to compare the results of Sections 
\ref{4.3} and \ref{4.4} with field theory, we consider the one-loop 
Feynman amplitude describing the radiative correction to the propagator 
of a particle of square energy $p^2$, in a theory with a three-point 
vertex. For simplicity, we consider spinless particles. The starting 
point is the Euclidean expression
\be
A &=& \int \dd^D k~ \frac{1}{k^2 + m^2_a} ~ \frac{1}{(p-k)^2 + m^2_b}  \\
&=& \int_0^\infty\int_0^\infty \dd a \dd b ~\int \dd^D k~ \e^{-a(k^2 + m^2_a) -b\big((p-k)^2 + m^2_b\big)} \\
&=& \pi^{D/2} \int_0^\infty \frac{\dd \tau_2}{\tau_2^{D/2 -1}} \int_0^1 
\dd y~ \e^{Ny(1-y)\tau_2 - y \tau_2 m^2_a - (1-y)\tau_2 m^2_b} 
\ee
We have made here the continuation to the Minkowski space $p^2 = -N$. 
Now, in the corner $\tau_2 \longrightarrow \infty, y \longrightarrow 0$  (cf.~Eq.~(\ref{pinch}))
the leading contribution comes from the massless case $m_b^2 =0$, and
\be
A &=&  \pi^{D/2} \int_0^\infty \frac{\dd \tau_2}{\tau_2^{D/2 -1}} \int_0^1 
\dd y~  \e^{-Ny^2\tau_2 + (N-m^2_a) y \tau_2} \label{star} ~.
\ee
The dominant contribution comes from $y=(N-m^2_a)/(2N)$, therefore $y=0$
corresponds to $n/N \longrightarrow 0$ (taking $m^2_a = N-n$).

Moreover, let us now expand $y=0+\eta$ and compare the result with Eq.~(\ref{4.17}). 
Remember that
$\sqrt{Q}/\tau_2 \longrightarrow 1$ and that
$|\delta\phi| = (\Im\zeta)^2$. Then taking $\Im\zeta=\eta\tau_2$ we see
\be
t^N &\sim& \e^{-2\pi N \eta^2 \tau_2}~.
\ee
Since $\tau_2 \sim \ln N$ the relevant range is $\eta \sim 1/\sqrt{N}$ neglecting $\ln N$ corrections.
Furthermore, in Eq.~(\ref{4.17}) there is no linear term in $\eta$ in the exponent; it matches Eq.~(\ref{star}) 
if $n/\sqrt{N} \longrightarrow 0$. This implies that the energy of the massless particle is soft.

\end{document}